\documentclass[11pt,twoside]{article}

\usepackage{asp2006}

\begin{document}
\title{Study of ISM tracers in galaxies}   
\author{V. Casasola$^{1,2}$, L. Piovan$^{1,3}$, G. Galletta$^{1}$, D. Bettoni$^{4}$, E. Merlin$^{1}$}   
\affil{$^{1}$ Astronomy Department, Vicolo dell'Osservatorio 2, I-35122 Padova\\
$^{2}$ Paris Observatory-LERMA, 61 Av. de l'Observatoire, F-75014 Paris\\
$^{3}$ Max-Planck-Institut f\"ur Astrophysik, Garching bei M\"unchen, Germany\\
$^{4}$ INAF-OAPd, Vicolo dell'Osservatorio 5, I-35122 Padova}    

\begin{abstract} 
We collected data for two samples of normal and interacting galaxies for a total
of 2953 galaxies having fluxes in one or more of the following wavebands: 
FIR, 21 cm line, CO(1-0) lines and soft X-ray. The large set of data obtained
allowed us to revisit some of the already known relations between the different
tracers of the interstellar medium (ISM), such as the link between the FIR flux
and the CO line emission, the relation between X-ray emission and the blue or FIR
luminosity. The relation lacking from observations for early-type galaxies
has been discussed and explained in detail in the frame of a suitable theoretical model,
obtained by coupling chemo-dynamical N-body simulations with a dusty spectrophotometric
code of population synthesis.
\end{abstract}



\section{Introduction}   
Some relationships between the various phases of the interstellar gas and between gas,
dust, and stars have been studied since many years, such as the one between CO and
far-infrared (FIR); on the other side some relations, connected with X-ray emission, 
have been studied only recently. Using data presented in the literature and our
catalogues of normal and interacting galaxies \citep{normal, interacting}, 
we have collected data on fluxes in the 60, 100 $\mu$m, CO(1-0) lines, and in the soft 
X-ray band. This set of data allowed us to study and extend the relations 
between different tracers of the ISM in galaxies of various morphological types 
(from E to Irr), with different type of interaction (perturbed or normal galaxy) and 
activity (Seyfert 1, Seyfert 2, or Liners), as described in the following.

\section{Cold gas and warm dust}
Several authors \citep{mirabel,solomon, bregman} found that the
global galaxy luminosity derived from CO(1-0) line is directly related with
the flux at 100 $\mu$m. The large set of data we used allowed us to more
clearly redefine this relationship, and in particular we found
that the relation, first obtained by \citet{bregman} for early type
galaxies, is also valid for late type galaxies. This link derives
from the excitation of gas clouds by the currently forming stars and by the
warming of the dust in the galaxy.

\section{X-ray component}
It is known that a proportionality exists between L$_X$ produced by
discrete sources and L$_B$, the blue luminosity of the whole galaxy:
late-type galaxies have a global X-ray luminosity directly proportional to L$_B$,
while early-type systems are dominated by emission produced by hot diffuse gas, and
their $L_X$ is proportional to the square power of the blue luminosity, as
discussed by \citet{beuing}. For this reason, we studied the early and late-type
galaxies separately.

\subsubsection{Late-type galaxies}
In late-type galaxies (t$>$Sb) our data show the existence of a
linear relation between soft X-ray fluxes and other indicators of
recent and current star formation (B and FIR luminosity respectively).
\citet{fabbiano0} have interpreted the link between B and X-ray luminosity
in late type galaxies as due to the contribution of discrete X-ray sources,
whose number is proportional to the quantity of already formed stars.
The X-ray emission is also produced by HII regions, where there is an ongoing
vigorous star formation, and its contribution appears more
evident in FIR light and may explain the existence of the
linear relation between L$_X$ and L$_{FIR}$.

\subsubsection{Early-type galaxies}
Considering the relations involving X-ray emission for
the early-type galaxies, the previous correlations become less evident,
in particular between L$_X$ and L$_{FIR}$.
To understand this apparent disagreement, we
used recent chemo-dynamical models (see \citealt{piovan})
coupled with dusty evolutionary population synthesis.
The luminosities derived by the models seem to confirm our
hypothesis about a connection between the exhaustion of the star
formation and the ``migration'' of the early type galaxies toward lower L$_{FIR}$, 
moving the galaxy representative points above the linear relation in the 
L$_X$ vs L$_{FIR}$ diagram. In most of early-type galaxies of our sample, 
the mechanism of IR emission is no longer strictly related to
the ongoing star formation and to the reprocessing of the radiation in the
dense regions where new stars are born. The FIR emission therefore comes
most likely from circumstellar dusty shells around AGB stars and
from an interstellar diffuse medium due to the outflow of dusty gas
from AGB and RGB stars.


\acknowledgements 
We would like to thank the LOC at ``From Stars to Galaxies: Building the pieces
to build up the Universe'' Conference (Venice, 0ctober 16-20, 2006) for supporting
one of us (VC).


\end{document}